\documentclass[conference]{IEEEtran}
\IEEEoverridecommandlockouts
\ifCLASSINFOpdf
  \usepackage[pdftex]{graphicx}
  \DeclareGraphicsExtensions{.pdf,.jpeg,.png}
\else
\fi

\usepackage{cite}
\usepackage{amsmath,amssymb,amsfonts}
\usepackage{bm}
\usepackage{comment}
\usepackage{algorithm, algorithmic}
\usepackage{mwe}
\usepackage{siunitx}
\usepackage{svg}
\usepackage[]{graphicx}
\graphicspath{{figures/}}
\usepackage[nolist]{acronym}
\usepackage{url}

\DeclareSIUnit{\sample}{S}
\DeclareSIUnit{\bit}{b}

\title{Uplink Multi-User MIMO Testbed Implementation in OpenAirInterface
    \thanks{
        The authors acknowledge the financial support by the Federal Ministry for Research, Technology and Space (BMFTR) in Germany in the programme of ``Souver\"an. Digital. Vernetzt.,'' joint project xG-RIC, project identification numbers: 16KIS2429K and 16KIS243, and by the 6G-MIRAI project, which has received funding from the Smart Networks and Services Joint Undertaking (SNS JU) under the European Union's Horizon Europe research and innovation program under Grant Agreement No 10119236.
        Views and opinions expressed are however those of the author(s) only, and they do not necessarily reflect those of the European Union or the SNS JU (granting authority). Neither the European Union nor the granting authority can be held responsible for them.
        
        The authors thank Ralf Lindstedt, Angelo Athanassopoulos, and Andreas Schiller of Fraunhofer \ac{HHI} for their valuable technical assistance during the building and testing of this system.
    }
}

\author{
    \IEEEauthorblockN{        
        Utku Uçak\IEEEauthorrefmark{1},
        Fariba Armandoust\IEEEauthorrefmark{1},
        Matthias Mehlhose\IEEEauthorrefmark{1},
        Daniel Schäufele\IEEEauthorrefmark{1},\\
        Jochen Fink\IEEEauthorrefmark{1},
        Renato L. G. Cavalcante\IEEEauthorrefmark{1},
        S{\l}awomir Sta\'nczak\IEEEauthorrefmark{1}\IEEEauthorrefmark{2},
    }
    \IEEEauthorblockA{\IEEEauthorrefmark{1} \textit{Fraunhofer Heinrich Hertz Institute}, Berlin, Germany}
    \IEEEauthorblockA{\IEEEauthorrefmark{2} \textit{Technische Universität Berlin}, Berlin, Germany}
}
\date{March 2025}

\begin{acronym}
  \acro{3GPP}{3rd Generation Partnership Project}
  \acro{5G}{5th Generation}
  \acro{Rx}{receive}
  \acro{FFT}{fast Fourier transform}
  \acro{IQ}{in-phase and quadrature}
  \acro{LO}{local oscillator}
  \acro{USB}{universal serial bus}
  \acro{DMRS}{demodulation reference signal}
  \acro{OFDM}{orthogonal frequency division multiplexing}
  \acro{FPGA}{field programmable gate array}
  \acro{HHI}{Heinrich Hertz Institute}
  \acro{UE}{user equipment}
  \acro{MIMO}{Multiple-Input Multiple-Output}
  \acro{MU-MIMO}{Multi-User MIMO}
  \acro{O-RAN}{Open Radio Access Network}
  \acro{OAI}{OpenAirInterface}
  \acro{SRS-RAN}{Software Radio Systems - Radio Access Network}
  \acro{PUSCH}{Physical Uplink Shared Channel}
  \acro{PUCCH}{Physical Uplink Control Channel}
  \acro{SRS}{Sounding Reference Signal}
  \acro{MRC}{Maximum Ratio Combining}
  \acro{RZF}{Regularized Zero Forcing}
  \acro{PRB}{Physical Resource Block}
  \acro{RU}{Radio Unit}
  \acro{DU}{Distributed Unit}
  \acro{CU}{Central Unit}
  \acro{GPS}{Global Positioning System}
  \acro{QPSK}{quadrature phase shift keying}
  \acro{PHY}{physical layer}
  \acro{MAC}{medium access control layer}
  \acro{USRP}{Universal Software Radio Peripheral}
  \acro{DSP}{digital signal processing}
  \acro{RRC}{radio resource control}
  \acro{PC}{personal computer}
  \acro{gNB}{next-Generation Node-B}
  \acro{TDD}{time domain duplexing}
  \acro{SFP}{small form-factor pluggable}
  \acro{SDR}{software defined radio}
  \acro{MCS}{modulation and coding scheme}
\end{acronym}

\newcommand{\paren}[1]{\left( #1 \right)}             

\newcommand{\complex}{\mathbb{C}}          

\newcommand{\channelmatrix}{\bm{H}}
\newcommand{\combinermatrix}{\bm{V}}
\newcommand{\identitymatrix}{\bm{I}}
\newcommand{\phasematrix}{\bm{\Theta}}
\newcommand{\zeromatrix}{\bm{0}}

\newcommand{\receivevector}{\bm{y}}
\newcommand{\noisevector}{\bm{n}}

\newcommand{\transmitvector}{\bm{x}}

\newcommand{\noisevariance}{\sigma}
\newcommand{\numusers}{K}

\newcommand{\numrxant}{M}

\newcommand{\idxuser}{k}
\newcommand{\phase}{\theta}
\newcommand{\delay}{\tau}
\newcommand{\regularization}{\lambda}

\newcommand{\estimate}[1]{\hat{#1}}
\newcommand{\hermtrans}[1]{#1^{\mathsf{H}}}
\newcommand{\inverse}[1]{\paren{#1}^{-1}}

\newcommand{\complexnormal}{\mathcal{C}\mathcal{N}}

\DeclareMathOperator{\diag}{diag}

\begin{document}

\bstctlcite{bstctl:nodash}

\maketitle

\begin{abstract}
    Cell-Free \ac{MIMO} and \ac{O-RAN} have been active research topics in the wireless communication community in recent years.
    As an open-source software implementation of the \ac{3GPP} \ac{5G} protocol stack, \ac{OAI} has become a valuable tool for deploying and testing new ideas in wireless communication systems.
    In this paper, we present our \ac{OAI}-based real-time uplink \ac{MU-MIMO} testbed developed at Fraunhofer \ac{HHI}.
    As a part of our Cell-Free \ac{MIMO} testbed development, we built a 2x2 \ac{MU-MIMO} system using general purpose computers and commercially available \acp{SDR}.
    Using a modified \ac{OAI} \ac{gNB} and two unmodified \ac{OAI} \ac{UE}, we show that it is feasible to use \ac{SRS} channel estimates to compute uplink combiners.
    Our results verify that this method can be used to separate and decode signals from two users transmitting in non-orthogonal time-frequency resources.
    This work serves as an important verification step to build a complete Cell-Free \ac{MU-MIMO} system that leverages \ac{TDD} reciprocity to perform downlink beamforming over multiple cells.
\end{abstract}

\acresetall

\section{Introduction}

The idea of using multiple antennas to exploit spatial diversity brought about the \ac{MIMO} revolution.
Enabling multiple cells to dynamically cooperate to serve locations that would otherwise be at the cell edge led to the concept of cell-free networks.
Meanwhile, the effort to reduce vendor lock-in by developing network equipment with open specifications and interfaces, led to the emergence of \ac{O-RAN}.
Thanks to open-source software projects that implement the \ac{3GPP} RAN protocol stack, it is becoming possible to build wireless networks for research and deployment, using general-purpose computers and \acp{SDR}, controlled by open-source firmware. 

In the last two decades, many \ac{MIMO} demos and proof-of-concept systems have been published in the literature.
Some of the first real-time demonstrators were built at the Fraunhofer \ac{HHI} \cite{hausteinMIMOOFDMCellularDeployment2007}.
These systems used \acp{FPGA} and \ac{DSP} boards to handle strict computational requirements.
Massive \ac{MIMO} systems with up to 100 antennas were reported in \cite{shepardArgosPracticalManyantenna2012,malkowskyWorldsFirstRealTime2017, vieiraFlexible100antennaTestbed2014}.
Those systems also required custom boards and high-performance \acp{FPGA} to handle stringent signal processing tasks.
Testbeds and demonstrators using \ac{MIMO} in mmWave frequencies were built in \cite{blandinoMultiUserFrequencySelectiveHybrid2018, haqueM3MIMO8x8MmWave2024}.
Although these systems proved the feasibility of \ac{MIMO} hardware, they focused on \ac{PHY} with a custom waveform and protocol, as scalability and interoperability were not their focus.

Two main open-source projects that implement the \ac{3GPP} protocol stack for research and deployment are \ac{OAI} \cite{oai2014} and \ac{SRS-RAN} \cite{srsrandevelopmentteamSrsRAN2025}.
An \ac{O-RAN}  cell-free demonstrator was recently implemented using \ac{SRS-RAN} \cite{chuTestbedDevelopmentIntelligent2025}.
\ac{OAI} supports basic \ac{O-RAN} functionality like \ac{DU}-\ac{CU} split, Split~7.2 \cite{ORANWG4CUS0v1601} with commercial \acp{RU}, and Split~8 \cite{ORANWG7IPCHRDOpt8v0100} to stream time-domain \ac{IQ} samples to a \ac{USRP}.
\ac{OAI} also supports single-user \ac{MIMO} as described in \cite{saaifanNRMIMOFeature2021}.
In \cite{buiImplementationMultiUserMIMO2025}, the authors implemented downlink \ac{MU-MIMO} in \ac{OAI}; however, to our knowledge, there is no project that adds uplink \ac{MU-MIMO} functionality or cell-free support to \ac{OAI}.
We identified this feature as an essential capability for demonstrating the full potential of cell-free \ac{O-RAN} systems.
The defining capabilities of cell-free networks such as cooperative, phase-coherent multi-antenna transmission and reception from distributed locations, and beamforming using only uplink pilots by exploiting the \ac{TDD} channel reciprocity require \ac{MU-MIMO} as a prerequisite feature.
To close this gap, we describe our real-time uplink \ac{MU-MIMO} implementation on \ac{OAI} in this paper.
This system is a key component of the cell-free testbed we are developing at Fraunhofer \ac{HHI}.

The remainder of the paper is organized as follows.
Section~\ref{sec:system-description} gives a general overview of the system parameters and functions.
In Section~\ref{sec:theoretical_background}, we review the relevant theoretical concepts.
Sections~\ref{sec:phy_implementation}~and~\ref{sec:mac_implementation} detail the \ac{PHY} and \ac{MAC} implementation, respectively.
In Section~\ref{sec:hardware} we elaborate on the hardware we used, and in Section~\ref{sec:results} we discuss our results.

\section{System Description} \label{sec:system-description}

\begin{figure*}[!ht]
  \centering
  \includegraphics[width=\linewidth,keepaspectratio]{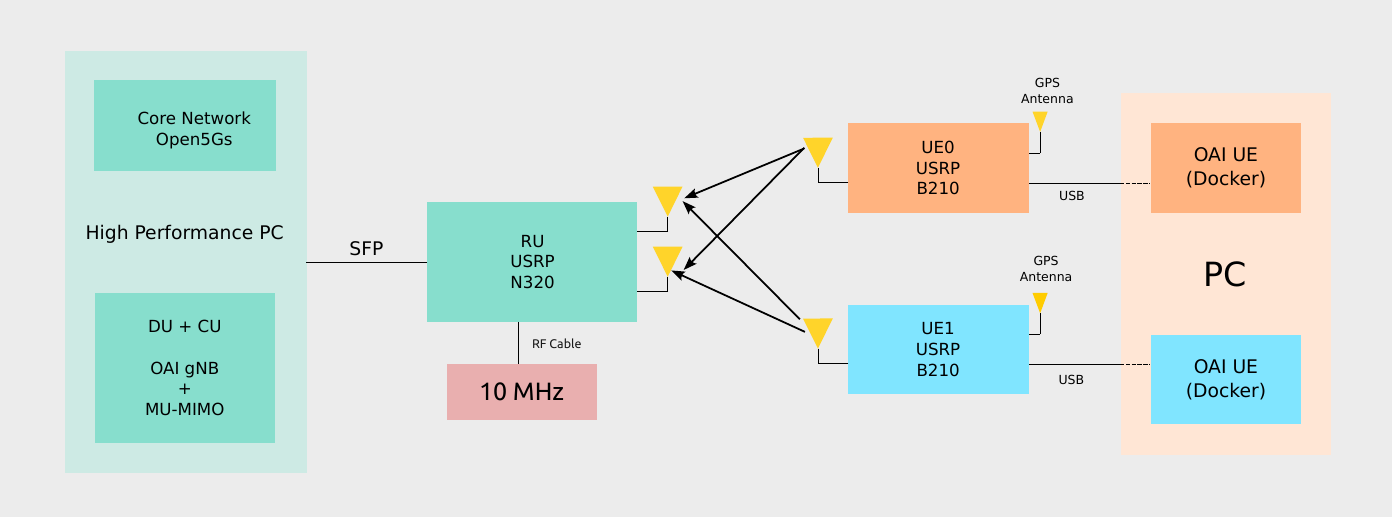}
  \caption{Block diagram of main system components.}
  \label{fig:system-diagram}
\end{figure*}

As shown in the block diagram in Fig.~\ref{fig:system-diagram}, our system consists of a monolithic \ac{gNB} with two antennas and two users, each with a single antenna, so we have a 2x2 \ac{MU-MIMO} system.
The complex baseband signal is based on \ac{3GPP}'s Release 15 specification, which is an \ac{OFDM} signal with cyclic prefix.
The total system bandwidth is \SI{10}{\mega\hertz} (24 \acp{PRB}).
As \ac{3GPP} defines a \ac{PRB} as 12 consecutive subcarriers, there are 288 subcarriers to encode symbols in each \ac{OFDM} symbol, which requires a 512-point \ac{FFT} for \ac{OFDM} processing.

The waveform uses the numerology 1 defined in \cite{3GPPTS38211v1620}, which corresponds to \SI{30}{\kilo\hertz} subcarrier spacing.
Given the \ac{FFT} size, the overall system sampling rate is equal to \SI{15.36}{\mega\sample\per\second}.
We operated the system at a carrier frequency of \SI{3319.68}{\mega\hertz}, which belongs to band n78.
These parameters are summarized in Table~\ref{tab:waveform-params}.

The uplink physical layer signal processing loop with nonorthogonal data scheduling can be summarized as follows: 
Following the cell search and random access procedures, the network configures each user to periodically transmit pilot symbols.
These orthogonal pilots span the bandwidth used for nonorthogonal data transmission.
The \ac{gNB} uses the received pilots to estimate the \ac{MIMO} channel and compute a linear combiner matrix.

When users need to transmit data, the scheduler assigns time and frequency resources in a separate layer to each user.
This means that two users are given permission to transmit in the same time-frequency resources in a nonorthogonal fashion, causing interference to the other user.
Once the \ac{gNB} receives the uplink signal with multi-user interference using two antennas, it applies the \ac{MU-MIMO} combiner to the received signal vector.
This operation has the objective of separating the symbols of both users.
From this point, the recovered data symbols are processed in the same way as the received symbols from a single antenna system with orthogonal scheduling.


\begin{table}[ht]
  \centering
  \caption{Waveform Parameters}
  \label{tab:waveform-params}
  \begin{tabular}{|l|c|}
    \hline
    \textbf{Config Name} & \textbf{10 MHz}  \\ \hline
    No. Resource Blocks  & 24  \\ \hline
    No. Subcarriers  & 288 \\ \hline
    \ac{FFT} Size & 512  \\ \hline
    Sampling Rate & \SI{15.36}{\mega\sample\per\second} \\ \hline
    Subcarrier Spacing & \SI{30}{\kilo\hertz} \\ \hline
    Carrier Frequency & \SI{3319.68}{\mega\hertz} \\ \hline
    \acs{MCS} & 4 \\ \hline
    Modulation & \acs{QPSK} \\ \hline
    Code rate & $1/2$ \\ \hline
  \end{tabular}
\end{table}

\section{Theoretical Background} \label{sec:theoretical_background}

In this section, we briefly review the essential theory behind the linear \ac{MIMO} detection and the effect of residual frequency offset to the detection process.

\subsection{MU-MIMO and Linear Detection}

The \ac{MIMO} system model with a flat-fading channel with $\numusers$ transmit and $\numrxant$ receive antennas is given by

\begin{equation*}
    \receivevector = \channelmatrix \transmitvector + \noisevector;
\end{equation*}
where $\receivevector \in \complex^\numrxant$ is the signal received from the \ac{Rx} array, $\transmitvector \in \complex^\numusers$ is the transmit signal vector, $\channelmatrix \in \complex^{\numrxant \times \numusers}$ is the \ac{MIMO} channel matrix, and $\noisevector \sim \complexnormal  \paren{\zeromatrix, \noisevariance^2\identitymatrix_\numrxant}$ is the additive white Gaussian noise vector.
Each variable corresponds to one \ac{OFDM} subcarrier.
The subcarrier indices are dropped for the sake of brevity.
For the case of \ac{MU-MIMO}, $\transmitvector$ is a column vector where each element is the transmit symbol of a separate user.
The receiver needs to estimate $\transmitvector$ by applying a linear operation to $\receivevector$ as

\begin{equation}
    \estimate{\transmitvector} = \hermtrans{\combinermatrix} \receivevector,
    \label{eq:mimo_reception}
\end{equation}
where $\estimate{\transmitvector}$ is the estimated transmit vector and $\hermtrans{\combinermatrix} \in \complex^{\numusers \times \numrxant}$ is the \ac{MIMO} combiner matrix.
We consider two common \ac{MIMO} detection algorithms; namely, \ac{MRC} and \ac{RZF}, $\hermtrans{\combinermatrix}_{\mathrm{MRC}} = \hermtrans{\channelmatrix}$, and $\hermtrans{\combinermatrix}_{\mathrm{RZF}}=\inverse{\hermtrans{\channelmatrix} \channelmatrix + \regularization\identitymatrix} \hermtrans{\channelmatrix}$, where $\identitymatrix$ is the identity matrix and $\regularization$, a nonnegative real number, is the regularization parameter.

\subsection{Two Step Channel Equalization} \label{sec:two_step_channel_equalization}

The model in (\ref{eq:mimo_reception}) assumes that the receiver has perfect channel state information.
The residual carrier frequency offset between the transmitter and the receiver becomes an important factor when periodic pilots independent of the user data are used to calculate $\combinermatrix$.
Due to the time difference $\delay_\idxuser$ between the periodic pilot transmission of the user $\idxuser$ and the data symbol transmission, the channel vector of each user undergoes a phase rotation $\phase_\idxuser$, proportional to the frequency offset and $\tau_\idxuser$.
Here we assume that the channel has a coherence time longer than the maximum $\delay_\idxuser$, and the channel has not changed much except the phase rotation.
We can represent the phase rotation with the diagonal matrix $\phasematrix = \diag(e^{j\phase_1}, \ldots, e^{j\phase_\numusers})$ from the individual phases.
As $\channelmatrix$ represents the channel state in the instance of channel estimation, the model during data transmission becomes

\begin{equation*}
    \receivevector = \channelmatrix \phasematrix \transmitvector + \noisevector.
\end{equation*}

Applying \ac{RZF} gives us

\begin{equation*}
    \hermtrans{\combinermatrix} \receivevector = \phasematrix \transmitvector + \Tilde{\noisevector}
\end{equation*}
which is the user signal vector with canceled interference, but with additional phase rotation.
The noise vector is transformed as $\Tilde{\noisevector} = \hermtrans{\combinermatrix}\noisevector \sim \complexnormal \paren{\zeromatrix, \noisevariance^2\hermtrans{\combinermatrix}\combinermatrix}$.
The remaining phase rotation can be estimated using a single-layer channel estimation from the \ac{DMRS} pilots embedded inside the data symbols and equalized by rotating the symbols back accordingly.

\section{Physical Layer Implementation} \label{sec:phy_implementation}

Most of the modification necessary to implement uplink \ac{MU-MIMO} in \ac{OAI} was in \ac{gNB}'s \ac{PHY} \ac{Rx} function. In this section, we explain the features that were necessary to implement uplink \ac{MU-MIMO}.  

\begin{figure*}[htbp]
  \centering
  \includegraphics[width=\linewidth,keepaspectratio]{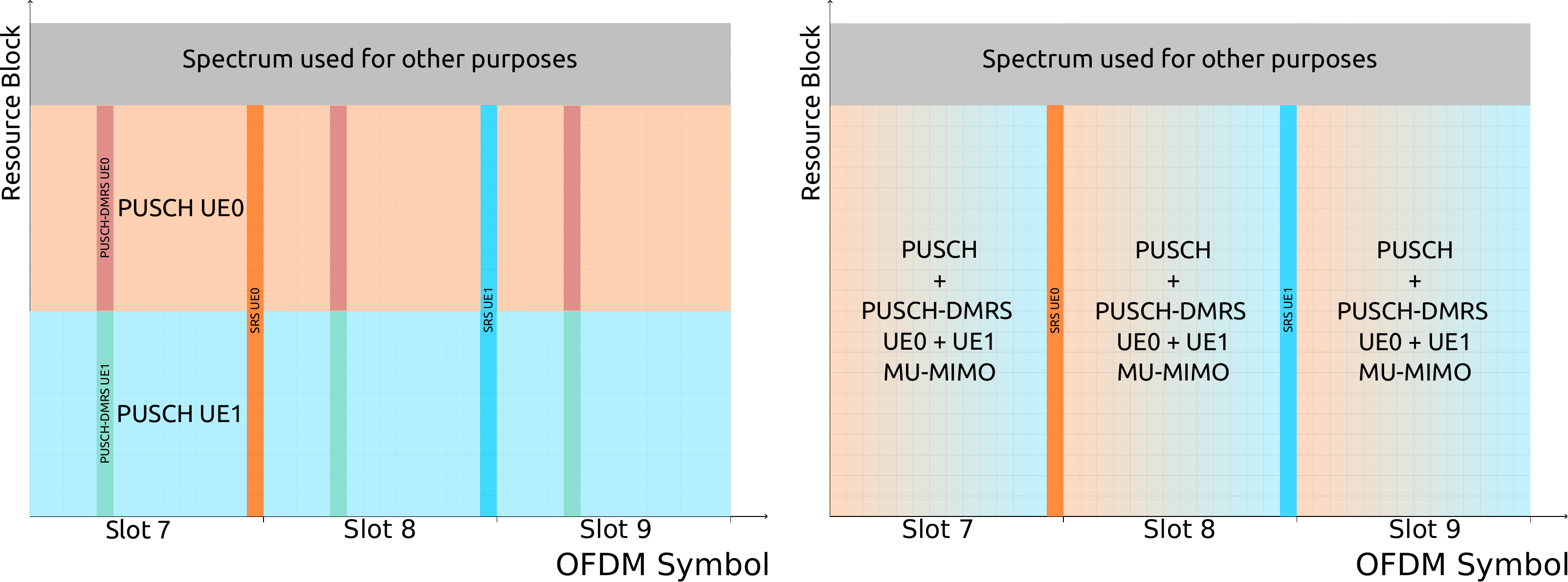}
  \caption{A simplified \ac{OFDM} grid showing three uplink slots with physical channel allocation when \ac{MU-MIMO} scheduling is disabled (left) and enabled (right).}
  \label{fig:ofdm-grid}
\end{figure*}

\subsection{Open Source RAN Stack: Open Air Interface}

The software implementation of this project is built as a private feature branch of the Open Air Interface \cite{oai2014} project.
Our implementation is based on the \texttt{2024.w45} version of the official repository.
Before detailing the \ac{PHY} implementation of the \ac{MU-MIMO} processing, we will provide details on how \ac{OAI} handles \ac{PHY} \ac{Rx} processing.
As summarized in Algorithm~\ref{alg:vanilla_oai_ul_rx}, \ac{PHY} functions for the \ac{gNB} receiver run in their own thread.
The main worker function of this thread, \texttt{phy\_procedures\_gNB\_uespec\_RX} is called for each uplink slot and processes every physical channel scheduled to that slot.
It is implemented as three consecutive for-loops.
The first loop iterates over all \ac{PUCCH} instances expected to arrive in that slot.
The same operation is performed for \ac{PUSCH}, which carries the user data.
Lastly, \ac{SRS} is processed, whose function will be detailed in the next section.

The \texttt{DecodePUSCH} procedure performs channel estimation and equalization using \ac{PUSCH} \ac{DMRS}, soft quadrature-amplitude-demodulation, descrambling, and channel decoding to recover \ac{MAC} data payload.
Similar operations are performed for the control channel in \texttt{DecodePUCCH} routine.
As the name suggests, the \texttt{SRSChannelEstimation} routine estimates the wireless channel from \ac{SRS} pilots.

\subsection{\acl{SRS}}

The \ac{SRS} is a physical channel designed to transmit pilot symbols from users to the base station to perform uplink channel measurements.
It can be configured to be sent periodically from each user in a wide bandwidth.
\ac{SRS} plays an important role for our system because we use it to estimate channel state information to compute the \ac{MU-MIMO} combiner ($\channelmatrix$ and $\hermtrans{\combinermatrix}$ from Section~\ref{sec:theoretical_background}).

There are several options to multiplex \ac{SRS} from multiple users.
In this project, we use a purely TDD approach for simplicity of implementation.
Each user is scheduled to transmit a wideband pilot sequence for one \ac{OFDM} symbol in a different slot.
The pilots are transmitted twice per radio frame (every \SI{5}{\milli\second}) and each user transmits their pilots in adjacent slots.
Thi results in a time offset of approximately \SI{.5}{\milli\second} between the channel measurement instances of the first and second user.

\begin{algorithm}[!hb]
\caption{\ac{OAI} \ac{gNB} \ac{Rx} \ac{PHY} Processing (simplified)}
\label{alg:vanilla_oai_ul_rx}
\begin{algorithmic}[0]
  \REQUIRE frame\_idx, slot\_idx, PUCCH\_set, PUSCH\_set, 
  \STATE SRS\_set
  \FORALL{PUCCH \textbf{in} PUCCH\_set}
    \STATE DecodePUCCH(PUCCH)
  \ENDFOR
  \FORALL{PUSCH \textbf{in} PUSCH\_set}
    \STATE DecodePUSCH(PUSCH)
  \ENDFOR
  \FORALL{SRS \textbf{in} SRS\_set}
    \STATE SRSChannelEstimation(SRS)
  \ENDFOR
\end{algorithmic}
\end{algorithm}

As we now use \ac{SRS} channel estimates for data decoding, it is important to use fresh channel state information as soon as it is available.
To achieve this, we reversed the order of the \ac{PUSCH} and \ac{SRS} loops in the receiver function.
As soon as new channel information arrives, the \ac{RZF} combiner matrix is recomputed with up-to-date channel state information.

\subsection{PUSCH Processing}

We implemented the multi-user interference suppression as a pre-processing step to the regular \ac{PUSCH} decoding routine.
Once the \ac{MU-MIMO} combiner is applied to the receive vector, the rest of the processing can be performed as if the signal were coming from a single antenna without interference.
The modified \ac{PHY} \ac{Rx} algorithm is given in Algorithm~\ref{alg:oai_ul_rx_mu_mimo}.
With this approach, the \ac{OAI} codebase for \ac{PUSCH} processing remains mostly unmodified, while adding \ac{MU-MIMO} interference cancellation capability.
As \ac{DMRS} channel estimation and equalization are part of this process, the phase rotation explained in Section~\ref{sec:two_step_channel_equalization} is corrected.

\begin{algorithm}[!hb]
\caption{Physical layer receiver function with \ac{MU-MIMO} Support (simplified)}
\label{alg:oai_ul_rx_mu_mimo}
\begin{algorithmic}[0]
\REQUIRE frame\_idx, slot\_idx, PUCCH\_set, PUSCH\_set,
\STATE SRS\_set

\FORALL{PUCCH in PUCCH\_set}
    \STATE DecodePUCCH(PUCCH)
\ENDFOR

\FORALL{SRS in SRS\_set}
    \STATE $\channelmatrix \gets$ SRSChannelEstimation(SRS)
    \STATE $\combinermatrix \gets$ ComputeRZFCombiner($\channelmatrix$)
\ENDFOR

\FORALL{PUSCH in PUSCH\_set}
    \IF{is\_mu\_mimo\_active}
        \STATE MUMIMOPreprocessing(PUSCH, $\combinermatrix$)
    \ENDIF
    \STATE DecodePUSCH(PUSCH)
\ENDFOR

\end{algorithmic}
\end{algorithm}

\subsection{Why SRS Instead of DMRS?}

The design choice of using \ac{SRS} channel estimates for computing the combiner instead of \ac{PUSCH}-\ac{DMRS} can be explained as follows:
The next milestone for this system is to perform downlink \ac{MU-MIMO} precoding by leveraging the uplink-downlink channel reciprocity in a TDD system.
A parallel project is underway to develop a calibration step to equalize the mismatch between the transmit and receive radio-frequency frontends.
In order to achieve this, periodic updates on the channel state information are necessary to keep the downlink precoder matrix up-to-date.
This would not be possible if the uplink pilots were only sent when there is uplink data to transmit.
Our work shows that \ac{SRS} can be configured and used to decode data symbols on the uplink.
This was a necessary validation step before using \ac{SRS}-based channel estimates to compute the downlink precoders.

\subsection{Fixed vs Floating Point Arithmetic}

The \ac{DSP} operations of \ac{OAI}'s \ac{PHY} use fixed-point arithmetic.
The reason for this design choice is real-time performance, as processors perform integer operations faster than floating-point operations.
Fixed-point operations can also be parallelized more efficiently with single-input-multiple-data operations.
However, implementing complex algorithms with integer arithmetic is cumbersome.
The designer must ensure that all the signal levels are appropriately scaled to minimize overflows and underflows.
Inverting a matrix is especially challenging in this case, as division by a small determinant might easily cause an overflow.
Because of that, we implemented our functions using floating point arithmetic.
After converting the \ac{SRS} channel estimates and the received \ac{IQ} samples to floating point representation, we computed the \ac{RZF} combiner matrix and applied it to the received signal.
We finally converted the equalized signal back to integers as \ac{OAI}'s \ac{PUSCH} processing function expects an integer signal as input.

Our experiments with network benchmarking tools showed that this approach was fast enough to keep up with the real-time processing constraints.
However, it is possible to further optimize this routine to balance speed and numerical accuracy. 
The idea behind this strategy lies in the following observation:
The combiner is computed only when new channel state information arrives, while every received data symbol must be equalized before it can be decoded.
This means that the combiner matrix is used much more often than it is computed.
The strategy is to keep the combiner computation in float, but to perform matrix vector multiplication between the combiner matrix and the receive vector using integers.
In this way, the step that is prone to errors with integers remains float, but the step that uses most of the computer cycles is optimized with integers.
We leave this idea for the next step of development in future work.

\subsection{User Equipment and Frequency Synchronization}

In this project, we experimented with commercial \ac{UE} as well as \ac{OAI} software \ac{UE} with \acp{USRP}.
\ac{MU-MIMO} scheduling and the processing done by the \ac{gNB} is transparent to \acp{UE}.
From a \ac{UE}'s perspective, a \ac{RRC} command configures \ac{SRS} transmission and when the user has uplink data to transmit, a \ac{PUSCH} grant is issued to them in a certain time and frequency resource.
The \ac{UE} does not know if another user is also transmitting in the same set of resources.
As a result, there is no need to modify the \ac{UE} to enable uplink \ac{MU-MIMO}.

The main difference between a commercial \ac{UE} and \ac{OAI} \ac{UE} was the possibility of using a reference signal for frequency stability.
We observed that \ac{OAI} software \acp{UE} are more stable when their local oscillators were conditioned with an external \SI{10}{\mega\hertz} reference signal or with \ac{GPS}.
Commercial \acp{UE} do not have the possibility to accept an external reference signal for frequency synthesis.
Although they could also connect to the cell, they lost synchronization and disconnected when high traffic was sent.

\section{MAC Implementation} \label{sec:mac_implementation}

\subsection{SRS Scheduling}

\ac{SRS} is usually used to measure long-term channel statistics.
Because of that, its period is typically configured to be relatively long.
In \ac{OAI}, the default period of \ac{SRS} is 8 radio frames (\SI{80}{\milli\second}).
To track the channel for multi-user interference management, we modified the \ac{SRS} scheduler so that it configures \acp{UE} to transmit \ac{SRS} pilots twice per radio frame (\SI{5}{\milli\second} period).

There are several options to multiplex \ac{SRS} from multiple users; such as \ac{TDD}, frequency interlacing with a comb pattern, and a mixture of these two approaches.
For simplicity, we choose a \ac{TDD} approach where each user transmits \ac{SRS} in neighboring slots.

The bandwidth of \ac{SRS} is represented in terms of \acp{PRB}.
Due to the way \ac{RRC} signaling for \ac{SRS} configuration is standardized by \ac{3GPP}, it is usually not possible to configure \ac{SRS} to the exact system bandwidth.
For example, 20 \acp{PRB} is the maximum \ac{SRS} bandwidth in a 24 \ac{PRB} system.
The missing \acp{PRB} at the edge do not get pilots to compute \ac{MU-MIMO} combiners.
We deactivated the \acp{PRB} without \ac{SRS} during \ac{PUSCH} scheduling, so no data is scheduled for those \acp{PRB}.
A simplified representation of the uplink OFDM scheduling can be seen in Fig.~\ref{fig:ofdm-grid}.

\subsection{Nonorthogonal PUSCH Scheduling}

To enable nonorthogonal scheduling of \ac{PUSCH} to multiple users, we modified \ac{OAI}'s \ac{PUSCH} scheduler, which is implemented in the \path{pf_ul} function inside \path{gNB_scheduler_ulsch.c} file.
Instead of marking a \ac{PRB} as occupied when one user is scheduled to it, \ac{PRB} allocation is tracked separately for both \ac{MU-MIMO} layers.
This leads to reusing time-frequency resources when scheduling both users and increasing the link capacity.

\section{Hardware Setup} \label{sec:hardware}

\begin{figure}[ht]
  \centering
  \includegraphics[width=\linewidth,keepaspectratio]{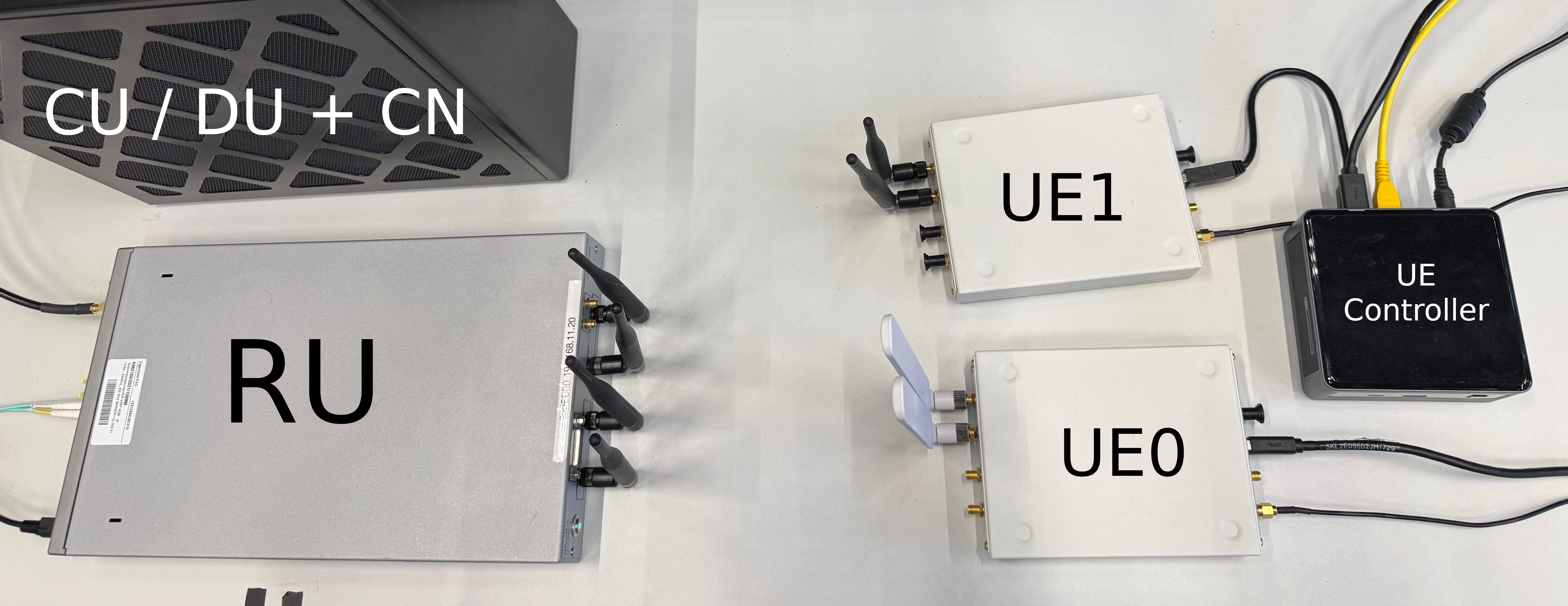}
  \caption{A picture of the setup in our lab.}
  \label{fig:setup_photo}
\end{figure}

Our \ac{MU-MIMO} system is made up of a \ac{gNB}, a core network, and two \acp{UE}.
We used commercially available \acp{SDR}, general-purpose computers, and free open-source software to build this system.
A modified version of \ac{OAI} \ac{gNB}, as detailed in Sections~\ref{sec:phy_implementation}~and~\ref{sec:mac_implementation} runs on a high-performance \ac{PC} with an Intel i9 processor.
We deployed a containerized Open5Gs instance on the same \ac{PC} as the core network.
We used a \ac{USRP} N320 with two antennas as the \ac{RU}.
The \ac{PC} and the \ac{USRP} exchange time domain \ac{IQ} samples over a \SI{10}{\giga\bit\per\second} Ethernet over fiber (\ac{SFP}) link.
A \SI{10}{\mega\hertz} reference signal is connected to the \ac{USRP} to stabilize its \ac{LO}.
A picture of our over-the-air setup can be seen in Fig.~\ref{fig:setup_photo}.

As \ac{UE}, we used two \ac{USRP} B210s with a single antenna controlled by \ac{OAI} \ac{UE} software running on a \ac{PC}.
We controlled two \acp{USRP} using a single \ac{PC} running two separate \ac{OAI} \ac{UE} instances in Docker containers.
To stabilize \ac{UE} \acp{LO}, we experimented with three methods: a \SI{10}{\mega\hertz} reference signal, \ac{GPS} disciplined oscillators, and relying on the frequency offset compensation mechanism inside \ac{OAI} \ac{UE}.
The first two methods worked well and resulted in a stable connection, while allowing the \acp{LO} to free-run while correcting only the baseband frequency in software caused frequent disconnections.
We also experimented with Quectel RM510 as a commercial \ac{UE}.
They showed a similar behavior to USRPs with a free-running \ac{LO}, suggesting that frequency synchronization is a critical factor to build stable uplink \ac{MU-MIMO} systems to preserve the orthogonality between subcarriers in the uplink signal.

\section{Results \& Discussion} \label{sec:results}

To validate our system, we visualized several \ac{PHY} signals and user throughput in real time as the system is running.
We are particularly interested in the \ac{SRS} channel estimates in frequency domain and \ac{PUSCH} symbols as they were received, \ac{MRC}, and \ac{RZF} equalized.
By changing the number of \ac{MU-MIMO} layers in the \ac{PUSCH} scheduler between one and two, we can enable and disable the \ac{MU-MIMO} functionality and observe the change in \ac{PUSCH} constellation and user throughput in real time.

\begin{figure}[ht]
  \centering
  \includegraphics[width=\linewidth,keepaspectratio]{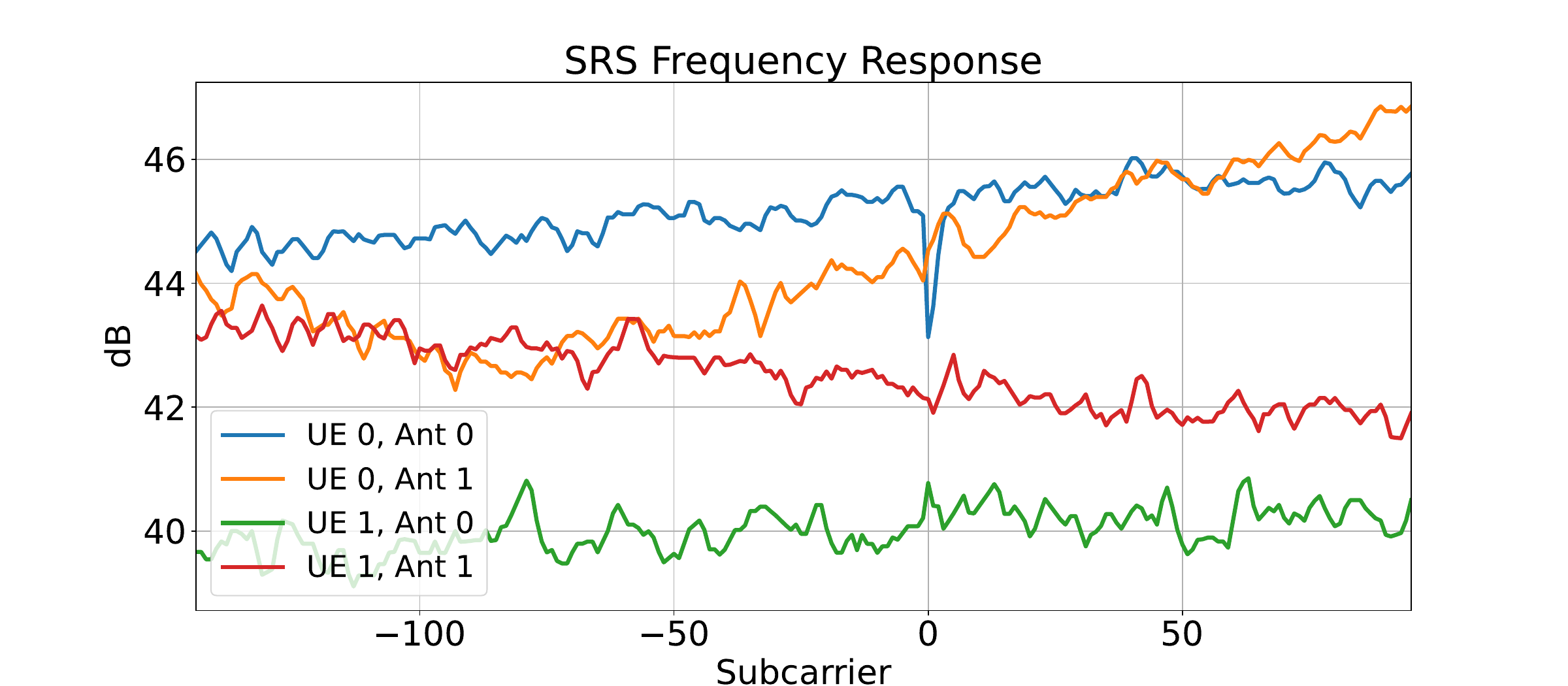}
  \caption{Channel frequency domain response from \ac{SRS} pilots.}
  \label{fig:srs_csi}
\end{figure}

In Fig.~\ref{fig:srs_csi} the magnitude frequency response of the \ac{MIMO} channel is visualized.
These measurements come from \ac{SRS} and are used to compute the \ac{MU-MIMO} combiner.
With a \SI{10}{\mega\hertz} bandwidth and a line-of-sight channel in our lab, we have a relatively flat channel.

\begin{figure}[htbp]
  \centering
  \includegraphics[width=\linewidth,keepaspectratio]{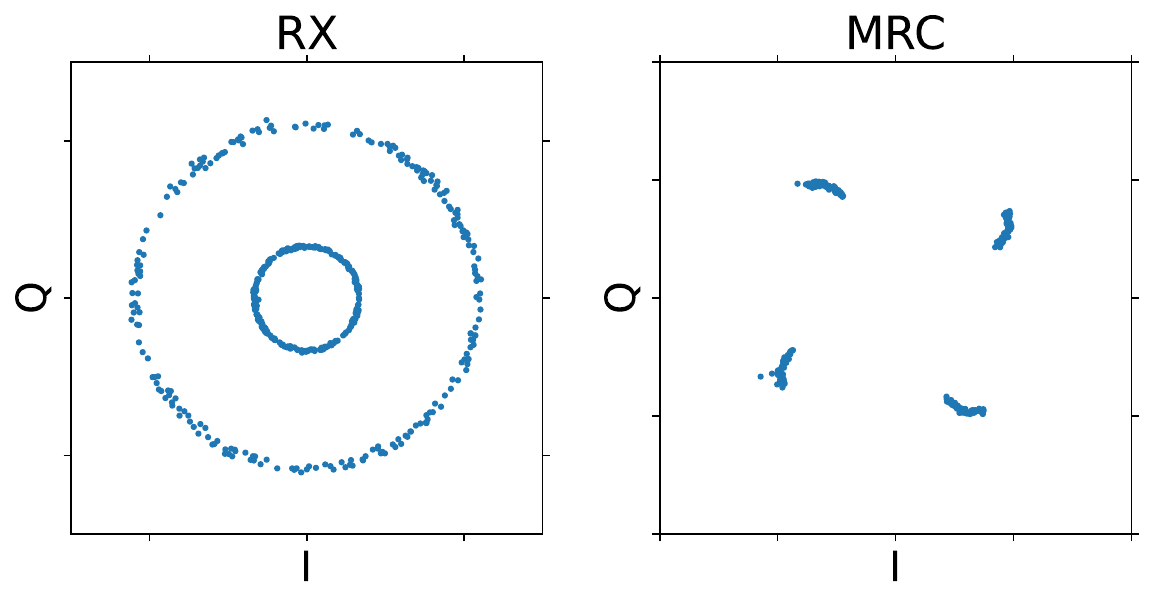}
  \caption{Received (left) and \ac{MRC} equalized (right) signal constellation of one of the users when \ac{MU-MIMO} scheduler is switched off.}
  \label{fig:constellation_mimo_off}
\end{figure}

\begin{figure}[ht]
  \centering
  \includegraphics[width=\linewidth,keepaspectratio]{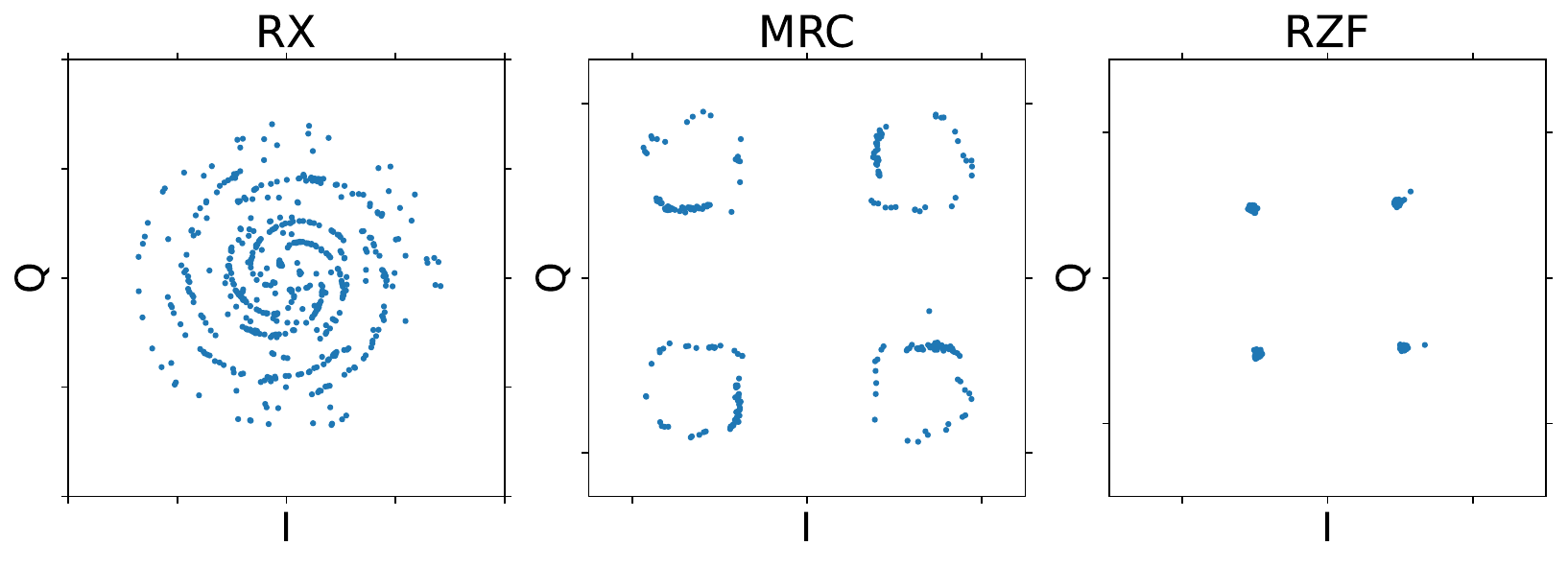}
  \caption{Received (left), \ac{MRC} equalized (center), and \ac{RZF} equalized (right) signal constellation of one of the users when \ac{MU-MIMO} scheduler is switched on.}
  \label{fig:constellation_mimo_on}
\end{figure}

As shown in Fig.~\ref{fig:constellation_mimo_off}, when \ac{MU-MIMO} scheduling is disabled, the received signal constellation is a circle.
This is the result of residual impairments, such as frequency offset.
Applying the \ac{MRC} combiner to this signal is sufficient to recover the \ac{QPSK} constellation.
In contrast, enabling \ac{MU-MIMO} scheduling (Fig.~\ref{fig:constellation_mimo_on}) introduces strong inter-user interference, which distorts the received symbols and prevents \ac{MRC} from producing a separable \ac{QPSK} pattern; and full channel inversion provided by the \ac{RZF} combiner is necessary to recover the constellation.

\begin{figure}[ht]
  \centering
  \includegraphics[width=\linewidth,keepaspectratio]{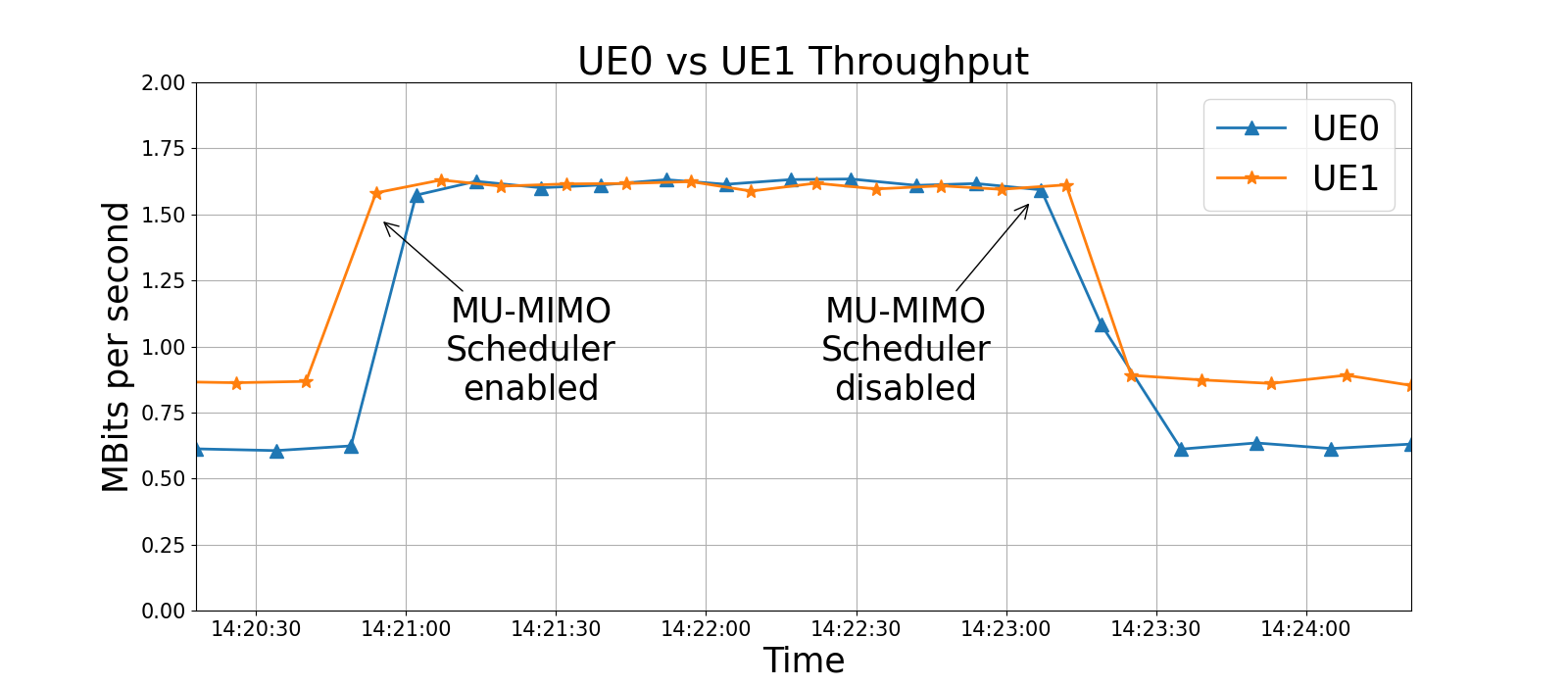}
  \caption{User throughput over time as the \ac{MU-MIMO} scheduler is switched on and off.}
  \label{fig:throughput}
\end{figure}

In Fig.~\ref{fig:throughput}, the effect on user throughput is clearly visible in the time series graph collected from throughput tests.
When the \ac{MU-MIMO} scheduler is enabled, the rate of each user is roughly doubled.
The increased rate is maintained as long as the \ac{MU-MIMO} scheduler remains enabled.
Both rates drop to their previous levels when the \ac{MU-MIMO} scheduler is switched off.

\section{Conclusion \& Future Work} \label{sec:conclusion}

In this paper, we introduced our real-time uplink \ac{MU-MIMO} system demo as a building block for our cell-free testbed.
Our results show that it is possible to build a \ac{MU-MIMO} system using general-purpose computers, commercial \acp{SDR}, and open source software.
In future work, we plan to extend this system with downlink \ac{MU-MIMO} capability using \ac{TDD} channel reciprocity with a hardware calibration step.
We also plan to scale the testbed by adding additional \acp{RU} in geographically separated locations, enabling a user-centric cell-free architecture.

\bibliographystyle{IEEEtran}
\bibliography{IEEEabrv,references}

\end{document}